\documentclass[a4paper,twocolumn,fleqn,superscriptaddress,preprintnumbers,prd]{revtex4}

\usepackage{graphicx,color}
\usepackage{amsmath,amssymb,amsfonts}

\newcommand{\be}{\begin{equation}}
\newcommand{\ee}{\end{equation}}

\newcommand{\bea}{\begin{eqnarray}}
\newcommand{\eea}{\end{eqnarray}}

\newcommand{\ba}{\begin{eqnarray}}
\newcommand{\ea}{\end{eqnarray}}

\def\be{\begin{equation}}
\def\ee{\end{equation}}
\def\beq{\begin{eqnarray}}
\def\eeq{\end{eqnarray}}

\begin{document}

\input amssym.def
\input amssym.tex

\title{Small black holes in global AdS spacetime}

\preprint{HIP-2015-27/TH, INT-PUB-15-046 }
\author{Niko Jokela}
\affiliation{Department of Physics and Helsinki Institute of Physics, P.O.~Box 64, FI-00014 University of Helsinki, Finland}
\author{Arttu P\"onni}
\affiliation{Department of Physics and Helsinki Institute of Physics, P.O.~Box 64, FI-00014 University of Helsinki, Finland}
\author{Aleksi Vuorinen}
\affiliation{Department of Physics and Helsinki Institute of Physics, P.O.~Box 64, FI-00014 University of Helsinki, Finland}

\begin{abstract}
We study finite temperature correlation functions and quasinormal modes in a strongly coupled conformal field theory holographically dual to a small black hole in global Anti-de Sitter spacetime. Upon variation of the black hole radius, our results smoothly interpolate between known limits corresponding to large black holes and thermal AdS space. This implies that the quantities are continuous functions of energy density in the microcanonical ensemble, thus smoothly connecting the deconfined and confined phases that are separated by a first order phase transition in the canonical description.
\end{abstract}

\maketitle

%%%%%%%%%%%%%%%%%%%%%%%%%%%%%%%%%%%%%%%%%%%
\section{Introduction}
%%%%%%%%%%%%%%%%%%%%%%%%%%%%%%%%%%%%%%%%%%%

When formulated in global Anti-de Sitter space instead of the Poincar\'e patch thereof, the AdS/CFT correspondence relates type IIB string theory living in AdS$_{d+1}$ spacetime to a supersymmetric conformal field theory ($\mathcal{N}=4$ Super Yang-Mills (SYM) theory for $d=4$) living in a $d-1$-dimensional sphere \cite{Witten:1998qj}. This version of the duality has been actively studied in the past, not least because of the interesting properties of field theories defined in compact spacetimes (see e.g.~\cite{Yamada:2006rx} and references therein). Analogously to the Poincar\'e patch, the black hole (BH) background in global AdS is dual to the field theory in thermal equilibrium. This time, the BH however turns out to be thermodynamically unstable in the canonical ensemble, if its radius is small enough, i.e.~$r_\text{h}<r_\text{min}\equiv\sqrt{\frac{d-2}{d}} L$, where $L$ is the curvature radius of the space and we have assumed $d\geq 3$.  The space surrounding the small BHs is approximately flat, and as the specific heat of the BH is negative, it evaporates away; as a result, the gravity solution becomes thermal (or pure) AdS with no BH in the bulk \cite{Hawking:1982dh,Aharony:1999ti,Horowitz:1999uv}. On the contrary, large BHs with $r_\text{h}\geq r_\text{min}$ are always thermodynamically stable, as are small black holes in the microcanonical ensemble (cf.~a detailed discussion of this issue in \cite{Aharony:2003sx}).

From the field theory point of view, it is interesting to study boundary observables in the black hole spacetime. For large black holes in global AdS space, there is indeed ample literature covering e.g.~the quasinormal mode (QNM) spectra of a variety of field theory operators; see e.g.~\cite{Berti:2009kk} and references therein. Generically, the eigenfrequencies $\omega$ of the QNMs are complex, with the real parts corresponding to the energies and the imaginary parts to the damping rates of the modes. The modes are typically sorted according to the magnitudes of ${\rm Im}\, \omega$ and labeled by integers (overtones) $n$. Other than for extreme cases, such as $n=0$ and $n\to\infty$, exact results are hard to obtain, and one often needs to resort to numerical methods. Nevertheless, several different analytical schemes have been developed that approximate numerical results for large BHs very accurately \cite{Berti:2009kk}.  Although small AdS BHs have received much less focus than their large counterparts, the so-called resonance method \cite{Berti:2009wx} has been shown to reproduce their QNM spectra to a good accuracy (see also \cite{Konoplya:2002zu} for numerical results).

While a lot of attention has been focused on the QNMs, the functional form of the corresponding two-point functions is a much less studied topic, and we are in fact unaware of any existing results for the small BH background. This is somewhat surprising in light of the fact that unlike in the case of thermal AdS space, studied in \cite{Giddings:2001ii}, the highly convenient Son-Starinets prescription \cite{Son:2002sd} is available as soon as the spacetime includes an event horizon, so that the correlators should be rather straightforwardly available for both small and large BH backgrounds. For small BHs, this observation is most likely linked with the small number of studies of the nature and applications of their field theory duals that would go beyond establishing that in the microcanonical ensemble they describe the field theory system for a range of energy densities. For two prominent exceptions to this, see however refs.~\cite{Asplund:2008xd,Basu:2010uz}. 

In the paper at hand, our goal is to fill in a gap in the literature related to small BHs in global AdS spacetime by studying selected dual field theory Green's functions in this gravitational background. We find the topic important and worth studying for several reasons. First, in the canonical ensemble the dual field theory is known to have a first order deconfinement phase transition at some critical temperature $T_\text{c}$, where several physical quantities exhibit discontinuous behavior. It should be very interesting to investigate what happens to the same observables in the microcanonical ensemble, where the small BH is stable and one can continuously decrease the energy density below the critical one while still residing in the BH phase. Secondly, studies of equilibration in strongly coupled field theories often reduce to solving dynamical problems in global AdS spacetime that typically involve gravitationally collapsing objects, such as thin shells \cite{Danielsson:1999zt} or pressureless dust \cite{Giddings:2001ii,Taanila:2015sda}  (for reviews, see e.g.~\cite{CasalderreySolana:2011us,Brambilla:2014jmp}). In these calculations complications often arise from the initial geometry not involving an event horizon, which is typically remedied by introducing a small black hole as a regulator \cite{Chesler:2010bi,Bantilan:2014sra}. The question of what the quantitative --- or even qualitative --- impact of changing the background geometry in such a way is has received minimal attention in the literature. Finally, a closely related theme is also the investigation of whether small perturbations of thermal AdS spacetime can lead to black hole formation \cite{Bizon:2011gg,Abajo-Arrastia:2014fma,Dimitrakopoulos:2014ada}. We hope that our work will shed some light on all of these issues, as well as find practical applications in future work on related subjects.

Our paper is organized at follows. First, we introduce the framework needed to study two-point functions of a field theory operator dual to a bulk scalar field in the background of a small black hole in global AdS. Then, we proceed to display and analyze our results for the corresponding QNM spectrum and spectral function, varying both the radius of the black hole and the mass of the bulk scalar. Here, our main goal will be to demonstrate that in the limit of $r_\text{h}\to 0$, our results --- derived with the Son-Starinets prescription --- reduce to those of thermal AdS space, obtainable from the earlier work of ref.~\cite{Giddings:2001ii} (cf.~similar observations in the context of D-branes \cite{Myers:2007we,Mateos:2007yp}). After this, we spend some time dwelling on the interpretation of our results on the field theory side, recalling what is known about the phase structure of strongly coupled large-$N_c$ ${\mathcal N}=4$ SYM in $S^3$. Finally, we conclude with a brief summary of the findings and their implications for studies of gravitational collapse in global AdS space.

%%%%%%%%%%%%%%%%%%%%%%%%%%%%%%%
\section{Setup}
%%%%%%%%%%%%%%%%%%%%%%%%%%%%%%%%%%%%%

As we work in global AdS spacetime containing a black hole, our metric has the usual AdS-Schwarzschild form
\begin{align}
   ds^2=-f(r) dt^2 + \frac{dr^2}{f(r)} + r^2 d\Omega_{d-1}^2,
\end{align}
in which the constant-$r$ spatial sections are $d-1$-spheres. The function $f(r)$ appearing here reads
\begin{align}
   f(r)=1 - \frac{C}{r^{d-2}} + \frac{r^2}{L^2},
\end{align}
where $C$ is a constant proportional to the BH mass. It is related to the horizon radius $r_\text{h}$ by
\begin{align}
   C=r_\text{h}^{d-2} \left( 1 + \frac{r_\text{h}^2}{L^2} \right), 
\end{align}
which also sets the Hawking temperature of the black hole via
\begin{align}\label{eq:T}
   T= \frac{d-2+d(r_\text{h}/L)^2}{4\pi r_\text{h}}.
\end{align}
In the rest of this paper, we will set the curvature radius of the space to unity, $L=1$.

The AdS-Schwarzschild black hole is known to be thermodynamically stable as long as  $r_\text{h}\geq r_\text{min}=\sqrt{\frac{d-2}{d}}$  (recall, though, that for $r_\text{h}<1$ the thermodynamically preferred phase in the canonical ensemble is thermal AdS). For smaller values of the horizon radius, $r_\text{h}<r_\text{min}$, the small BH has a negative specific heat and will thus be unstable.  Its lifetime is, however, very large compared to its energy, which means that we can approximately treat the small BH as stationary and furthermore assign a temperature to characterize it. Interestingly, it is even possible to identify the dual field theory states, which correspond to configurations far from equilibrium \cite{Asplund:2008xd}. For these reasons and with the purposes of our paper in mind, we will from this point onwards not be concerned with the instability, although it would constitute a rather interesting exercise of its own to check if the lowest quasinormal mode of a gauge field is tachyonic, as predicted by the Gubser-Mitra conjecture \cite{Gubser:2000mm,Buchel:2005nt}.

Consider now a massive scalar field in the above spacetime, and assume it to be homogeneous in the directions along the $d-1$ -sphere. The equation of motion, $(\Box-m^2)\phi(t,r)=0$, then simplifies in Fourier space to
\begin{align}
   \frac{1}{r^{d-1}}\partial_r \left( r^{d-1} \frac{\partial_r\phi}{f(r)} \right) - \left( m^2 + \omega^2 f(r) \right) \phi = 0, \label{eq:scalar_eom}
\end{align}
which serves as the starting point of our exercise, aimed at finding the spectral function and QNMs corresponding to the boundary CFT operator dual to this field. In this process, we use the Son-Starinets prescription \cite{Son:2002sd}, which amounts to imposing ingoing boundary conditions for the field at the horizon,
\begin{align}
   \phi(\omega,u\to 1) \sim (1-u)^{-i \xi}, \quad \xi \equiv \frac{r_\text{h} \omega}{d-2+d r_\text{h}^2}, \label{eq:ingoing_wave}
\end{align}
where we have changed the radial coordinate to $u\equiv r_\text{h}/r$. We may then read off the retarded correlator from the near boundary expansion of the field,
\begin{align}
   \phi(\omega,u\to 0) \sim \, &\mathcal{A}(\omega) u^{\Delta_-} (1+{\mathcal O}(u^2)) \nonumber \\
   & + \mathcal{B}(\omega) u^{\Delta_+}(1+{\mathcal O}(u^2)),
\end{align}
where $\Delta_\pm=\frac{d}{2}\pm \sqrt{\left(\frac{d}{2}\right)^2+m^2}\equiv \frac{d}{2}\pm \nu$ are the conformal weights of the operator. According to the prescription, the retarded correlator $G_R$ is proportional to the ratio of the coefficients $\mathcal{A}$ and $\mathcal{B}$
\begin{equation}
G_R(\mathbf{k}=0,\omega) \sim -\frac{\mathcal{B}(\omega)}{\mathcal{A}(\omega)}, 
\end{equation}
while the spectral function $\chi$ is nothing but the imaginary part of this quantity 
\begin{equation}
   \chi(\omega) = -2 \,\mathrm{Im}\, G_R(0,\omega).
\end{equation}
The quasinormal modes are finally solved (usually numerically) from Eq.~(\ref{eq:scalar_eom}): They correspond to those values of $\omega$, which simultaneously satisfy the ingoing wave boundary condition of Eq.~(\ref{eq:ingoing_wave}) and the Dirichlet boundary condition at the boundary, $\phi(\omega,u\to 0)=0$.

To aid an eventual comparison of our results with known limits, let us finally recall, how one can obtain the above spectral function in thermal AdS phase using the results of \cite{Giddings:2001ii}. There it is shown that the (retarded) boundary correlator of interest can be given in terms of the normal modes of the scalar field in the form of Eq.~(3.34) of this reference. Fourier transforming the equation to momentum space and taking the imaginary part of the result in the limit of vanishing wave vector readily leads to the result
\begin{equation}
 \chi(\omega)\sim \sum_{n=0}^\infty \frac{\kappa_{n0}^2}{\omega_{n0}}\delta(\omega-\omega_{n0}), \label{GNres}
\end{equation}
where the quantities $\omega_{n0}$ and $\kappa_{n0}$ read \cite{Giddings:1999jq}
\begin{eqnarray}
\omega_{n0}&=&2n+\Delta_+, \label{eq:omn0}\\ 
\kappa_{n0}^2&=&\frac{2\omega_{n0}\Gamma(n+\Delta_+)\Gamma(n+\nu+1)}{\Gamma(n+1)\Gamma(n+d/2)\Gamma(\nu+1)^2} \label{eq:kn0}.
\end{eqnarray}
For the case $d=4$ and $m=0$, studied numerically in the next section (see also \cite{Burgess:1984ti}), these results simplify to
\begin{eqnarray}
\omega_{n0}&=&2(n+2),\\
 \frac{\kappa_{n0}^2}{\omega_{n0}} &=&\frac{1}{2}(n+1)(n+2)^2(n+3). \label{eq:anres}
\end{eqnarray}
It is worth stressing that to leading order in the large-$N_c$ limit, these results hold in the entire thermal AdS phase, irrespective of the value of the temperature.

%%%%%%%%%%%%%%%%%%%%%%%%%%%%%%%%%%%%%%%%%%%%%%%%%%%%%%%%%%%%%%%%%%
\begin{figure}[t]
\begin{center}
\includegraphics[width=8cm]{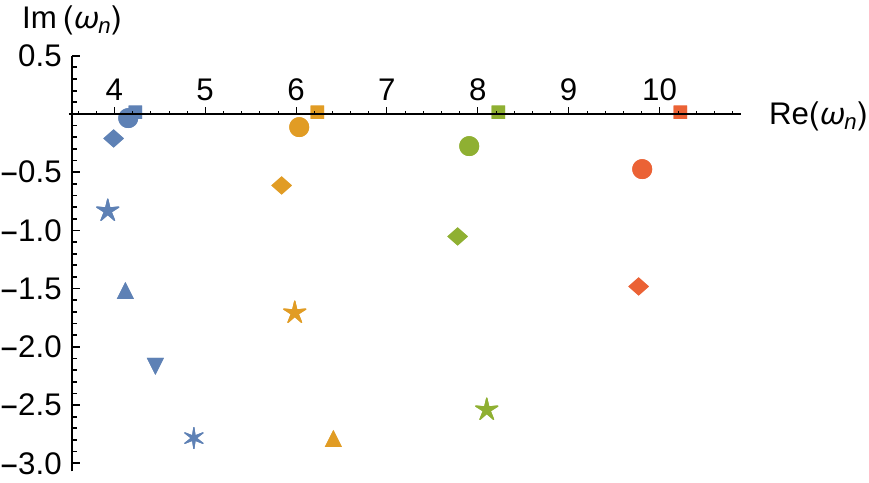}
\caption{The flow of the first four QNMs, as $r_\text{h}$ is lowered towards zero. The different colors stand for the different QNMs, while the different shapes correspond to $r_\text{h} = 0.0001,0.1,0.2,0.4,0.6,0.8,1$ (top to bottom). The mass of the field is set to unity here.}\label{fig:qnmflow}
\end{center}
\end{figure}
%%%%%%%%%%%%%%%%%%%%%%%%%%%%%%%%%%%%%%%%%%%%%%%%%%%%%%%%%%%%%%%%%%

%%%%%%%%%%%%%%%%%%%%%%%%%%%%%%%%%%%%%%%%
\section{Results}
%%%%%%%%%%%%%%%%%%%%%%%%%%%%%%%%%%%%%%%%

Having introduced our setup, we now proceed to inspect the results of the numerical determination of the QNMs and spectral functions, with the purpose of comparing them with the analytic low-temperature limit quoted above. To this end, recall that in addition to the spacetime dimensionality $d$ of the dual field theory, there are only two parameters in our setup, the mass $m$ of the scalar field and the horizon radius $r_\text{h}$, both of which we express in units of the AdS radius $L$. To the extent we have been able to check, there are no indications of any qualitative differences between different dimensions, so we will in the following only show results for $d=4$, unless otherwise stated. Our main focus will be in the transition towards the limit $r_\text{h}\to 0$, where we will compare our results to eqs.~(\ref{GNres})--(\ref{eq:kn0}).

%%%%%%%%%%%%%%%%%%%%%%%%%%%%%%%%%%%%%%%%%%%%%%%%%
\begin{figure}[t]
\begin{center}
\includegraphics[width=7cm]{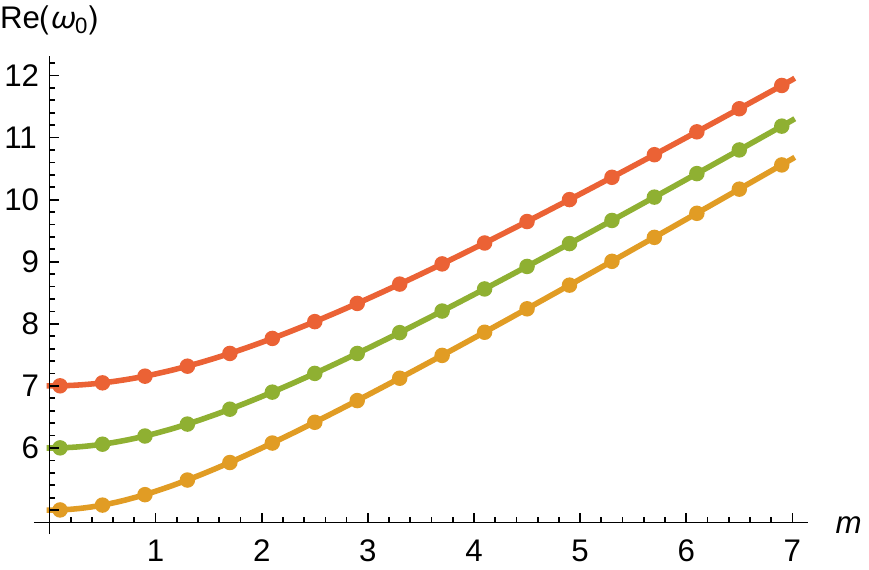}
\caption{The real part of the $n=0$ QNM frequency for $d=3,4,5$ (bottom to top). The dots correspond to our numerical results obtained for a discrete set of masses and $r_\text{h}=0.0001$, while the continuous curves are given by Eq.~(\ref{eq:omn0}).}\label{fig:qnm_vs_mass}
\end{center}
\end{figure}
%%%%%%%%%%%%%%%%%%%%%%%%%%%%%%%%%%%%%%%%%%%%%%%%%

To begin, we first illustrate in Fig.~\ref{fig:qnmflow} that the QNMs (or at least the first few overtones $n$) smoothly transition into normal modes when $r_\text{h}$ is lowered towards zero. We have explicitly checked that the imaginary parts of the modes vanish as $r_\text{h}^{d-2}$ for all values of the mass $m$, thus generalizing the $m=0$ results of \cite{Konoplya:2002zu}. As to the real part, we find non-monotonous behavior as a function of $r_\text{h}$ for all $m$ and $n$, again in accordance with the findings of \cite{Konoplya:2002zu}. Starting from largish values of $r_\text{h}$, the real part always first decreases with decreasing $r_\text{h}$, then reaches a minimum value, and finally begins to increase, asymptotically matching Eq.~(\ref{eq:omn0}). We have displayed the mass dependence of Eq.~(\ref{eq:omn0}) against our numerical results in Fig.~\ref{fig:qnm_vs_mass}, finding perfect agreement. 

%%%%%%%%%%%%%%%%%%%%%%%%%%%%%%%%%%%%%%%%%%%%
\begin{figure*}[t]
\begin{center}
\includegraphics[width=\textwidth]{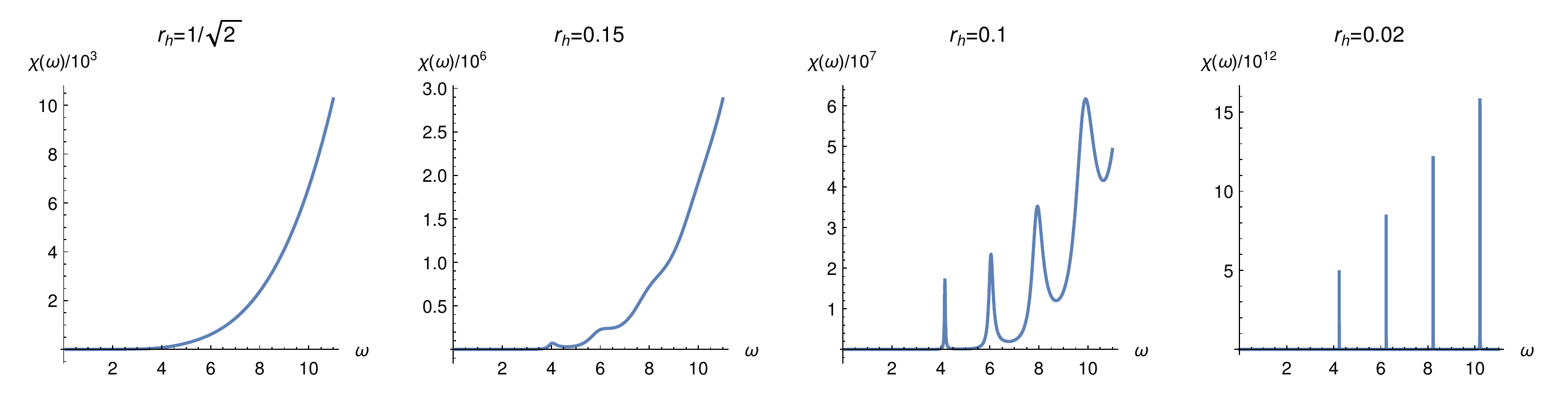}
\caption{The spectral function $\chi(\omega)$ for four different values of $r_\text{h}$, of which the first one corresponds to a marginally stable BH, $r_\text{h}=r_\text{min}$. The result is seen to continuously deform from a smooth and monotonous function towards a delta-comb distribution as $r_\text{h}\to 0$. The mass of the scalar field has again been set to unity here, and the normalization factor $N_c^2 T^2/4$ is omitted from the result.}\label{fig:spikes}
\end{center}
\end{figure*}
%%%%%%%%%%%%%%%%%%%%%%%%%%%%%%%%%%%%%%%%%%%

At finite values of $r_\text{h}$, the QNM spectra of the system are expected to be continuous in the sense that in appropriate channels there exists a hydrodynamical pole on the imaginary axis that tends to zero like $k^2$ in the limit of vanishing wavevector $k$. Exactly at $r_\text{h}=0$, there is on the other hand no BH in the bulk, and the spectrum should thus be discrete. These facts are in accordance with our observation that the QNMs lose their imaginary parts and thus become normal modes, as $r_\text{h}\to 0$. Further evidence for this transition can be extracted from the related spectral functions $\chi(\omega)$, which we analyze in Fig.~\ref{fig:spikes}. For finite values of $r_\text{h}$, $\chi(\omega)$ is seen to be a smooth, continuous function of the frequency $\omega$, which however starts to exhibit discrete bumps of increasing height, as one decreases $r_\text{h}$. In the limit $r_\text{h}\to 0$, the function finally turns into a delta-comb distribution, supported exactly at the frequencies $\omega_{n0}$ of Eq.~(\ref{eq:omn0}). This transition is illustrated by the four panels of Fig.~\ref{fig:spikes}, corresponding to four decreasing values of $r_\text{h}$.

The narrowing of the distinct `bumps', or resonances, of the spectral function in the limit of small $r_\text{h}$ and the simultaneous vanishing of the damping rates of the QNMs both point towards the system becoming describable in terms of long-lived quasiparticle degrees of freedom. It turns out that for small values of $r_\text{h}$, the full spectral function can to a very good accuracy be expressed as a sum over Lorentzian functions (damped harmonic oscillators) centered around the QNM poles,
\begin{equation}\label{eq:Lorentz}
 \chi(\omega)\sim \sum_{n=0}^\infty \frac{A_n}{(\omega-{\rm Re}\,\omega_{n})^2+({\rm Im}\,\omega_{n})^2}\, . 
\end{equation}
In the formal limit $r_\text{h}\to 0$, this function clearly approaches a sum of delta functions similar to Eq.~(\ref{GNres}), with only the coefficients $A_n$ to be determined.

\begin{figure*}[t]
\begin{center}
\includegraphics[width=0.7\textwidth]{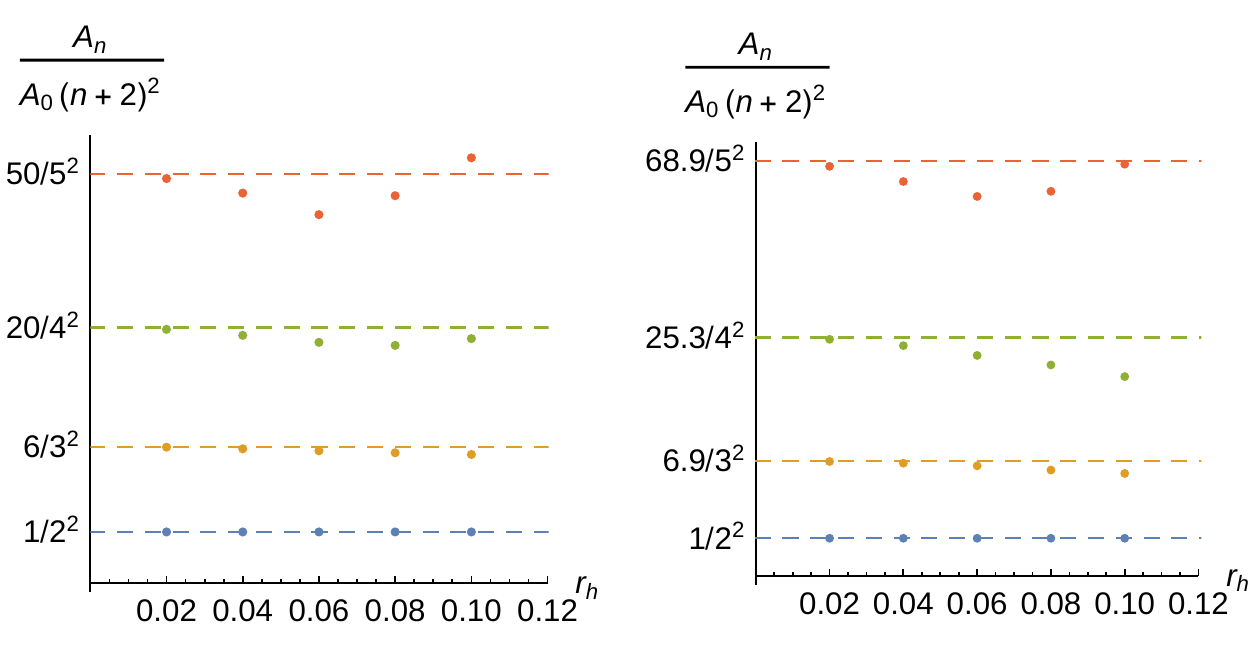}
\caption{The areas of the first four spikes of our spectral function, determined for several different values of $r_\text{h}$ and for $m=0$ (left) and $m=1$ (right), with the dashed lines corresponding to the coefficients of the delta functions in Eq.~(\ref{GNres}). For better visibility, the results have been normalized by $(n+2)^2$ times the area of the first spike.}\label{fig:areas}
\end{center}
\end{figure*}

To solve for the residues $A_n$ of Eq.~(\ref{eq:Lorentz}), we first note that they correspond to the areas below the bumps in the spectral function. To find their values, we perform the following analysis. We first construct an interpolating function from a set of local minima between the individual peaks, which we  then subtract from the spectral function. The resulting function consists of a set of well-defined separate bumps, whose areas we numerically evaluate to obtain the constants $A_n$, displayed for a few different values of $r_\text{h}$ in Fig.~\ref{fig:areas}. From here, we clearly see that in the limit of small $r_\text{h}$, the $A_n$'s indeed approach the analytical limit given by Eq.~(\ref{eq:anres}) for $d=4$ and $m=0$. The same behavior is observed for other values of $d$ and $m$, too, and it indeed appears that the spectral function approaches the thermal AdS limit of Eq.~(\ref{GNres}), when the size of the BH is taken to vanish. 

%%%%%%%%%%%%%%%%%%%%%%%%%%%%%%%%%%%%%%%%%%%%
\section{Discussion}
%%%%%%%%%%%%%%%%%%%%%%%%%%%%%%%%%%%%%%%%

To provide a proper interpretation of the above results on the field theory side, we will now specialize to $d=4$ and briefly recall what is known about the phase structure and thermodynamics of ${\mathcal N}=4$ SYM theory on $S^3$ to leading order in large $N_c$ and $\lambda$. We will do this separately for the canonical and microcanonical ensembles below, summarizing the results and discussion of \cite{Witten:1998qj,Aharony:2003sx}. 

In the canonical ensemble, the phase structure of the theory is most conveniently parameterized in terms of the temperature, given in units of the curvature radius of the AdS space. The temperature is a non-monotonous function of the BH radius, 
\begin{eqnarray}
 T(r_\text{h})&=&\frac{2r_\text{h}^2+1}{2\pi r_\text{h}}, \label{Trh}
\end{eqnarray}
which reaches a minimum at $r_\text{min}=1/\sqrt{2}$, corresponding to $T_\text{min} = \sqrt{2}/\pi$. Below this temperature, and in fact even between $T_\text{min}$ and $T_\text{c}=3/2\pi$ (corresponding to $r_\text{c}=1$), the  physical phase of the theory is given by the thermal AdS solution, i.e.~a gas of gravitons and other excitations in AdS space, which corresponds to the confined phase of the field theory. At $T=T_\text{c}$, the system exhibits a first order phase transition to the deconfined phase, described by the larger of the two black holes corresponding to the same temperature via Eq.~(\ref{Trh}). The smaller BH, studied in the previous sections, is never the preferred solution, but merely an unphysical, unstable saddle point of the functional integral. At the transition temperature, several physical quantities display discontinuous behavior, and e.g.~the energy density jumps from an ${\mathcal O}(N_c^0)$ value in the thermal AdS phase to a ${\mathcal O}(N_c^2)$ result in the BH phase.

%%%%%%%%%%%%%%%%%%%%%%%%%%%%%%%%%%%%%%%%%%%
%%%%%%%%%%%%%%%%%%%%%%%%%%%%%%%%%%%%%%%%%%%

In the microcanonical ensemble, the natural variable to parameterize the phase structure of the theory is the energy density $\epsilon$. Working to the leading order in large $N_c$ and $\lambda$, it can be shown that the thermodynamics of the system is always dominated by a black hole phase \cite{Aharony:2003sx}, with the relation between the energy density and the radius of the BH given by
\begin{eqnarray}
 \epsilon(r_\text{h})&=&\frac{3(r_\text{h}^2+r_\text{h}^4)}{16\pi G_\text{N}} = \frac{3(r_\text{h}^2+r_\text{h}^4)N_c^2}{8\pi^2} . \label{epsrh}
\end{eqnarray}
We can assign a statistical temperature to the system by differentiating the energy density with respect to the entropy density; the result, expressed in terms of the BH radius, then agrees with Eq.~(\ref{Trh}). As we can observe from these results, the shift from large to small BHs is not marked by a discontinuity in the energy density, but only by the specific heat of the system turning negative. 

\begin{figure*}[t]
\begin{center}
\includegraphics[width=1.0\textwidth]{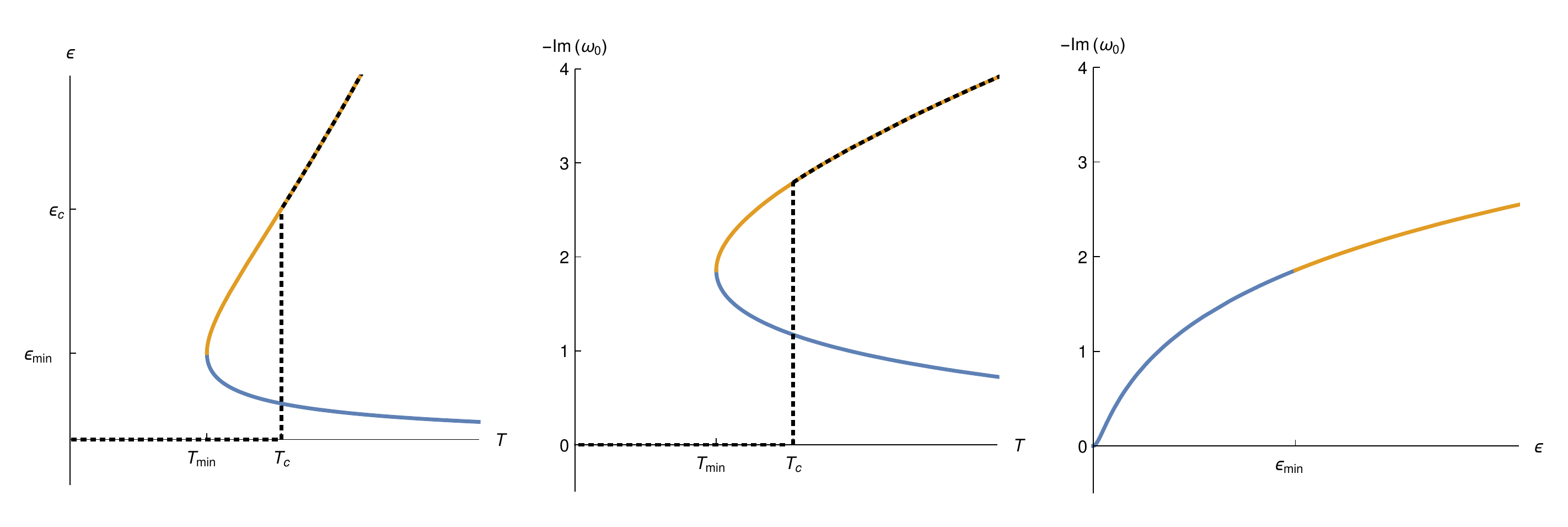}
\caption{Left and middle: The energy density and imaginary part of the first QNM as functions of temperature in the canonical ensemble. Right: The imaginary part of the first QNM as a function of energy density in the microcanonical ensemble. The subscript `c' refers to the critical temperature of the deconfinement transition, while `min' signifies the BH radius where the small and large BHs meet. The orange (blue) curves finally correspond to large (small) BHs, while the dashed black line indicates the physical behavior of the quantities in the canonical ensemble.}\label{fig:phases}
\end{center}
\end{figure*}

The above considerations are helpful in providing a field theory interpretation for the results we have obtained for small BHs in the previous sections. We see that the smooth interpolation of the QNMs and spectral function between the limits of large BHs and thermal AdS space represents the physical behavior of the system for a range of energy densities in the microcanonical ensemble, $\epsilon_0<\epsilon< 3N_c^2/(4\pi^2)$, where $\epsilon_0 \sim N_c^2/\lambda^{7/4}$ vanishes in the limit we are considering \cite{Aharony:2003sx}. The situation is illustrated in Fig.~\ref{fig:phases}, where we display the behavior of the energy density and the imaginary part of the first QNM $\omega_0$ in the canonical ensemble, as well as that of $-$Im($\omega_0$) in the microcanonical ensemble. It is worth stressing that the energy densities corresponding to small BHs in the microcanonical ensemble are not accessible in the canonical ensemble at all.

%%%%%%%%%%%%%%%%%%%%%%%%%%%%%%%%%%%%%%%%%%%
%%%%%%%%%%%%%%%%%%%%%%%%%%%%%%%%%%%%%%%%%%%

%%%%%%%%%%%%%%%%%%%%%%%%%%%%%%%%%%%%%%%%%%%%
\section{Conclusions}
%%%%%%%%%%%%%%%%%%%%%%%%%%%%%%%%%%%%%%%%

In the paper at hand, we have determined the quasinormal mode spectrum and the associated spectral function for a CFT operator dual to a massive bulk scalar field living in a global AdS$_{d+1}$-Schwarzschild background. We have concentrated on small black holes, seemingly irrelevant for the canonical ensemble of the dual field theory, and shown that in the limit of a pinching BH, $r_\text{h}\to 0$, the results approach those obtained earlier for thermal AdS$_{d+1}$. The observation has been given an interpretation in terms of the thermodynamics of the field theory in the microcanonical ensemble, which provides access to a range of energy densities not allowed in the canonical ensemble. 

For the QNMs the agreement of the small BH and thermal AdS results was conjectured already in \cite{Horowitz:1999jd}, but the perfect matching of the spectral functions --- i.e.~retarded correlators --- was not equally anticipated. This is in fact a very strong result, and prompts us to speculate that one might be able to smoothly match the complete sets of eigenfunctions in the small BH and thermal AdS cases, and thus continuously transition a non-Hermitian eigenvalue problem to a Hermitian one. The answer to whether this is the case appears to lie at the heart of understanding the black hole formation process in global AdS spacetime.

%%%%%%%%%%%%%%%%%%%%%%%%%%%%%%%%%%%%%%%%%%%%
\section{Acknowledgments}
%%%%%%%%%%%%%%%%%%%%%%%%%%%%%%%%%%%%%%%%

We thank P.~Chesler, C.~Hoyos, V.~Ker\"anen, A.~Karch, E.~Keski-Vakkuri, M.~Laine, M.~Lippert, P.~Romatschke, O.~Taanila, L.~Yaffe, and T.~Zingg for useful discussions, while A.P.~and A.V.~in addition acknowledge the Institute for Nuclear Theory in Seattle for its hospitality. Our work has been supported in part by the Academy of Finland grants no.~273545 and 1268023, as well as by the Magnus Ehrnrooth foundation.

%%%%%%%%%%%%%%%%%%%%%%%%%%%%%%%%%%%%%%%%%%%%%%%%%%%%%%


\begin{thebibliography}{99}


  
%\cite{Witten:1998qj}
\bibitem{Witten:1998qj}
  E.~Witten,
  %``Anti-de Sitter space and holography,''
  Adv.\ Theor.\ Math.\ Phys.\  {\bf 2} (1998) 253
  [hep-th/9802150];
  %%CITATION = HEP-TH/9802150;%%
  %\cite{Witten:1998zw}
%\bibitem{Witten:1998zw}
 % E.~Witten,
  %``Anti-de Sitter space, thermal phase transition, and confinement in gauge theories,''
  Adv.\ Theor.\ Math.\ Phys.\  {\bf 2} (1998) 505
  [hep-th/9803131].
  %%CITATION = HEP-TH/9803131;%%

%\cite{Yamada:2006rx}
\bibitem{Yamada:2006rx}
  D.~Yamada and L.~G.~Yaffe,
  %``Phase diagram of N=4 super-Yang-Mills theory with R-symmetry chemical potentials,''
  JHEP {\bf 0609} (2006) 027
  [hep-th/0602074].
  %%CITATION = HEP-TH/0602074;%%  
  
  %\cite{Hawking:1982dh}
\bibitem{Hawking:1982dh}
  S.~W.~Hawking and D.~N.~Page,
  %``Thermodynamics of Black Holes in anti-De Sitter Space,''
  Commun.\ Math.\ Phys.\  {\bf 87} (1983) 577.
  %%CITATION = CMPHA,87,577;%%
  
%\cite{Aharony:1999ti}
\bibitem{Aharony:1999ti}
  O.~Aharony, S.~S.~Gubser, J.~M.~Maldacena, H.~Ooguri and Y.~Oz,
  %``Large N field theories, string theory and gravity,''
  Phys.\ Rept.\  {\bf 323} (2000) 183
  [hep-th/9905111].
  %%CITATION = HEP-TH/9905111;%%

%\cite{Horowitz:1999uv}
\bibitem{Horowitz:1999uv}
  G.~T.~Horowitz,
  %``Comments on black holes in string theory,''
  Class.\ Quant.\ Grav.\  {\bf 17} (2000) 1107
  [hep-th/9910082].
  %%CITATION = HEP-TH/9910082;%%
  
  %\cite{Aharony:2003sx}
\bibitem{Aharony:2003sx}
  O.~Aharony, J.~Marsano, S.~Minwalla, K.~Papadodimas and M.~Van Raamsdonk,
  %``The Hagedorn - deconfinement phase transition in weakly coupled large N gauge theories,''
  Adv.\ Theor.\ Math.\ Phys.\  {\bf 8} (2004) 603
  [hep-th/0310285].
  %%CITATION = HEP-TH/0310285;%%
  %318 citations counted in INSPIRE as of 28 Aug 2015
  
%\cite{Berti:2009kk}
\bibitem{Berti:2009kk}
  E.~Berti, V.~Cardoso and A.~O.~Starinets,
  %``Quasinormal modes of black holes and black branes,''
  Class.\ Quant.\ Grav.\  {\bf 26} (2009) 163001
  [arXiv:0905.2975 [gr-qc]].
  %%CITATION = ARXIV:0905.2975;%%

%\cite{Berti:2009wx}
\bibitem{Berti:2009wx}
  E.~Berti, V.~Cardoso and P.~Pani,
  %``Breit-Wigner resonances and the quasinormal modes of anti-de Sitter black holes,''
  Phys.\ Rev.\ D {\bf 79} (2009) 101501
  [arXiv:0903.5311 [gr-qc]].
  %%CITATION = ARXIV:0903.5311;%%

%\cite{Konoplya:2002zu}
\bibitem{Konoplya:2002zu}
  R.~A.~Konoplya,
  %``On quasinormal modes of small Schwarzschild-anti-de Sitter black hole,''
  Phys.\ Rev.\ D {\bf 66} (2002) 044009
  [hep-th/0205142].
  %%CITATION = HEP-TH/0205142;%%
  
  

%\cite{Giddings:2001ii}
\bibitem{Giddings:2001ii}
  S.~B.~Giddings and A.~Nudelman,
  %``Gravitational collapse and its boundary description in AdS,''
  JHEP {\bf 0202} (2002) 003
  [hep-th/0112099].
  %%CITATION = HEP-TH/0112099;%%
  
%\cite{Son:2002sd}
\bibitem{Son:2002sd}
  D.~T.~Son and A.~O.~Starinets,
  %``Minkowski space correlators in AdS / CFT correspondence: Recipe and applications,''
  JHEP {\bf 0209} (2002) 042
  [hep-th/0205051].
  %%CITATION = HEP-TH/0205051;%%

 

%\cite{Asplund:2008xd}
\bibitem{Asplund:2008xd}
  C.~T.~Asplund and D.~Berenstein,
  %``Small AdS black holes from SYM,''
  Phys.\ Lett.\ B {\bf 673} (2009) 264
  [arXiv:0809.0712 [hep-th]].
  %%CITATION = ARXIV:0809.0712;%%
  
  %\cite{Basu:2010uz}
\bibitem{Basu:2010uz}
  P.~Basu, J.~Bhattacharya, S.~Bhattacharyya, R.~Loganayagam, S.~Minwalla and V.~Umesh,
  %``Small Hairy Black Holes in Global AdS Spacetime,''
  JHEP {\bf 1010} (2010) 045
  [arXiv:1003.3232 [hep-th]].
  %%CITATION = ARXIV:1003.3232;%%
  %34 citations counted in INSPIRE as of 28 Aug 2015
  
    
%\cite{Danielsson:1999zt}
\bibitem{Danielsson:1999zt}
  U.~H.~Danielsson, E.~Keski-Vakkuri and M.~Kruczenski,
  %``Spherically collapsing matter in AdS, holography, and shellons,''
  Nucl.\ Phys.\ B {\bf 563} (1999) 279
  [hep-th/9905227].
  %%CITATION = HEP-TH/9905227;%%
  
%\cite{Taanila:2015sda}
\bibitem{Taanila:2015sda}
  O.~Taanila,
  %``Holographic thermalization and Oppenheimer-Snyder collapse,''
  arXiv:1507.00878 [hep-th].
  %%CITATION = ARXIV:1507.00878;%%
  
  
  %\cite{CasalderreySolana:2011us}
\bibitem{CasalderreySolana:2011us}
  J.~Casalderrey-Solana, H.~Liu, D.~Mateos, K.~Rajagopal and U.~A.~Wiedemann,
  %``Gauge/String Duality, Hot QCD and Heavy Ion Collisions,''
  arXiv:1101.0618 [hep-th].
  %%CITATION = ARXIV:1101.0618;%%
  %320 citations counted in INSPIRE as of 04 août 2015sda
  
  %\cite{Brambilla:2014jmp}
\bibitem{Brambilla:2014jmp}
  N.~Brambilla {\it et al.},
  %``QCD and Strongly Coupled Gauge Theories: Challenges and Perspectives,''
  Eur.\ Phys.\ J.\ C {\bf 74} (2014) 10,  2981
  [arXiv:1404.3723 [hep-ph]].
  %%CITATION = ARXIV:1404.3723;%%
  %86 citations counted in INSPIRE as of 04 Aug 2015
 
 %\cite{Chesler:2010bi}
\bibitem{Chesler:2010bi}
  P.~M.~Chesler and L.~G.~Yaffe,
  %``Holography and colliding gravitational shock waves in asymptotically AdS_5 spacetime,''
  Phys.\ Rev.\ Lett.\  {\bf 106} (2011) 021601
  [arXiv:1011.3562 [hep-th]].
  %%CITATION = ARXIV:1011.3562;%%
  
  %\cite{Bantilan:2014sra}
\bibitem{Bantilan:2014sra}
  H.~Bantilan and P.~Romatschke,
  %``Simulation of Black Hole Collisions in Asymptotically Anti–de Sitter Spacetimes,''
  Phys.\ Rev.\ Lett.\  {\bf 114} (2015) 8,  081601
  [arXiv:1410.4799 [hep-th]].
  %%CITATION = ARXIV:1410.4799;%%
  
%\cite{Bizon:2011gg}
\bibitem{Bizon:2011gg}
  P.~Bizon and A.~Rostworowski,
  %``On weakly turbulent instability of anti-de Sitter space,''
  Phys.\ Rev.\ Lett.\  {\bf 107} (2011) 031102
  [arXiv:1104.3702 [gr-qc]].
  %%CITATION = ARXIV:1104.3702;%%
  
%\cite{Abajo-Arrastia:2014fma}
\bibitem{Abajo-Arrastia:2014fma}
  J.~Abajo-Arrastia, E.~da Silva, E.~Lopez, J.~Mas and A.~Serantes,
  %``Holographic Relaxation of Finite Size Isolated Quantum Systems,''
  JHEP {\bf 1405} (2014) 126
  [arXiv:1403.2632 [hep-th]].
  %%CITATION = ARXIV:1403.2632;%%
  
%\cite{Dimitrakopoulos:2014ada}
\bibitem{Dimitrakopoulos:2014ada}
  F.~V.~Dimitrakopoulos, B.~Freivogel, M.~Lippert and I.~S.~Yang,
  %``Instability corners in AdS space,''
  arXiv:1410.1880 [hep-th].
  %%CITATION = ARXIV:1410.1880;%%



  %\cite{Myers:2007we}
\bibitem{Myers:2007we}
  R.~C.~Myers, A.~O.~Starinets and R.~M.~Thomson,
  %``Holographic spectral functions and diffusion constants for fundamental matter,''
  JHEP {\bf 0711} (2007) 091
  [arXiv:0706.0162 [hep-th]].
  %%CITATION = ARXIV:0706.0162;%%
  
  %\cite{Mateos:2007yp}
\bibitem{Mateos:2007yp}
  D.~Mateos and L.~Patino,
  %``Bright branes for strongly coupled plasmas,''
  JHEP {\bf 0711} (2007) 025
  [arXiv:0709.2168 [hep-th]].
  %%CITATION = ARXIV:0709.2168;%%

  %\cite{Gubser:2000mm}
\bibitem{Gubser:2000mm}
  S.~S.~Gubser and I.~Mitra,
  %``The Evolution of unstable black holes in anti-de Sitter space,''
  JHEP {\bf 0108} (2001) 018
  [hep-th/0011127].
  %%CITATION = HEP-TH/0011127;%%
  %233 citations counted in INSPIRE as of 04 août 2015
  
%\cite{Buchel:2005nt}
\bibitem{Buchel:2005nt}
  A.~Buchel,
  %``A Holographic perspective on Gubser-Mitra conjecture,''
  Nucl.\ Phys.\ B {\bf 731} (2005) 109
  [hep-th/0507275].
  %%CITATION = HEP-TH/0507275;%%

  
  
  %\cite{Giddings:1999jq}
\bibitem{Giddings:1999jq}
  S.~B.~Giddings,
  %``Flat space scattering and bulk locality in the AdS / CFT correspondence,''
  Phys.\ Rev.\ D {\bf 61} (2000) 106008
  [hep-th/9907129].
  %%CITATION = HEP-TH/9907129;%%

%\cite{Burgess:1984ti}
\bibitem{Burgess:1984ti}
  C.~P.~Burgess and C.~A.~L\"utken,
  %``Propagators and Effective Potentials in Anti-de Sitter Space,''
  Phys.\ Lett.\ B {\bf 153} (1985) 137.
  %%CITATION = PHLTA,B153,137;%%
  
      %\cite{Horowitz:1999jd}
\bibitem{Horowitz:1999jd}
  G.~T.~Horowitz and V.~E.~Hubeny,
  %``Quasinormal modes of AdS black holes and the approach to thermal equilibrium,''
  Phys.\ Rev.\ D {\bf 62} (2000) 024027
  [hep-th/9909056].
  %%CITATION = HEP-TH/9909056;%%

  
  
\end{thebibliography}
\end{document}